\documentstyle{article}

\newcommand{\mathbf}{\bf}

\begin{document}

\begin{center}
{\huge\bf  The Edge Currents in IQHE }
\end{center}

\vspace{1cm}
\begin{center}
{\large\bf
F.GHABOUSSI}\\
\end{center}

\begin{center}
\begin{minipage}{8cm}
Department of Physics, University of Konstanz\\
P.O. Box 5560, D 78434 Konstanz, Germany\\
E-mail: ghabousi@kaluza.physik.uni-konstanz.de
\end{minipage}
\end{center}

\vspace{1cm}

\begin{center}
{\large{\bf Abstract}}
\end{center}

\begin{center}
\begin{minipage}{12cm}
It is shown that an observed length in the potential drops across  
IQHE samples  is a universal length for a given magnetic field  
strength which has the magnitude equal to the reciprocal magnitude  
of magnetic length and which results from the quantum mechanical  
uncertainty relation in presence of magnetic field. The analytic  
solution of Ohm's equation for the potential in Corbino sample in  
IQHE is also given.

\end{minipage}
\end{center}

\newpage

We recently showed that the microscopic theory of IQHE \cite{all}  
can be given by the canonical quantization of a semi-classical  
theory of the usual "classical" Hall-effect CHE \cite{mein}.

The action functional for this is the semi-classical  
Schroedinger-Chern-Simons one for a 2-D non-interacting carrier  
system with the usual minimal electromagnetic coupling on a   
2+1-dimensional manifold $M = \Sigma\times\mathbf R$ with spatial  
boundary
which results in the Ohm's equations as the equations of motion of  
the coupled electromagnetic potential. We showed also that the  
constraints of the theory under typical conditions of IQHE  
\cite{kk}, i. e. with small carrier concentration and higher  
magnetic field, forces the coupled electromagnetic potential to be  
an almost pure gauge potential and it forces the potential to exist  
only close to the boundary of $\Sigma$ \cite{mein}. Accordingly, the  
edge currents are the prefered currents under the IQHE conditions,  
in view of the mentioned constraints of the theory.

Here we show that the recent results on the potential drops across  
IQHE samples near the edges \cite{DK} folow the universal  
uncertainty relations of quantum mechanics as it should be to expect  
in view of the universality of the QHE.

Let us first explain from the more fundamental point of view of  
quantum mechanics.

\medskip
For charged systems, e. g. electrons in magnetic fields, the energy  
uncertainty is given by the minimum amount of the energy, i. e. the  
ground state energy. This amount of energy is proportional to the  
applied magnetic field strength. On the other hand, an energy  
uncertainty is correlated with a position uncertainty for the  
circulating electrons in magnetic field. Thus, quantum mechanically  
in presence of magnetic fields there is always an uncertainty of  
position of the electronic currents which is related with the width  
of the electron orbit.
Therefore, if we consider the uncertainty of momentum equal to $(2m  
\Delta E)^{\frac{1}{2}}$ with $\Delta E = E_{n+1} - E_{n} =  
{\displaystyle{\frac{\hbar \omega_c}{2}}}$ and $\omega_c =  
{\displaystyle{\frac{eB}{M_e}}}$, then the mentioned uncertainty is  
given by $\Delta X =  
({\displaystyle{\frac{\hbar}{eB}}})^{\frac{1}{2}}$ which is the  
magnetic length $l_B$. Since, the edge current is defined as the  
current which flows, in the ideal case, close to the edge within the  
length scale of the magnetic length \cite{kk}. This means that one  
should expect that according to  Ohm's equation for QHE, in the  
ideal case, also the potential distribution on the sample should be  
close to the boundary of sample within a distance which is  
proportional to the magnetic length. In view of the relations  
between the magnetic field strength $B$, magnetic length and the  
global density of elctrons $n$ with the filling factor $\nu$, i. e.  
$l^2_B = {\displaystyle{\frac{\hbar}{eB}}} =  
{\displaystyle{\frac{\nu}{2\pi n}}}$, it is obvious that a variation  
of only one of these factors changes the magnetic length and so it  
changes also the current position and the potential distribution on  
the sample. On the other hand, obviously if $B$ or  
${\displaystyle{\frac{\nu}{n}}}$ remain the same for various IQHE  
samples the magnetic length should be invariant for all these  
samples under the IQHE conditions independent of their geometries  
and other factors. This is the quantum theoretical basics of what is  
observed in the mentioned experiments for the potential drops  
\cite{DK}, where the authors report that they observed potential  
drops across the IQHE-samples over a length of $100 \mu m$ from the  
edge of samples. We show that this length which has the {\it  
magnitude} of $|l_B^{-1}|$ for the given data in Ref. \cite{DK} is  
indeed a universal quantity for a given $B$ or for a given  
${\displaystyle{\frac{\nu}{n}}}$ \cite {nn}.

\medskip
There is however one basic point with respect to the  
electromagnetic potential distribution which must be taken into  
account, namely that a potential is itself no observable in view of  
its gauge dependence. The observables related with the potential or  
those related with its field strength are phase angle given by the  
circle integral of potential and the surface integral of field  
strength which are observable by the quantum mechanical interfrence  
patterns. Equivalently, a constant potential multiplied by a proper  
length, e. g. by the circumference of mentioned line integral is  
also observable. For example according to the definition of magnetic  
length $l^{2}_B = {\displaystyle{\frac{\hbar}{e B}}}$ we have (see  
also belov):

\begin{equation}
l^{2}_B B = l_B A = \frac{\hbar}{e}\;\;\;,
\end{equation}
\label{m lang}

which is equivalent to the definition of magnetic flux quantum  
through $\int\int B = \oint A = {\displaystyle{\frac{h}{e}}}$, where  
$A = l_B. B$ is the relevant component of electromagnetic gauge  
potential in the $A_m = B.x_n \epsilon_{mn}$ gauge.

\medskip
Moreover as a general result let us mention that, if one considers  
the relation (3) in form $2 \pi l_B A = 2 \pi l^{2}_B B =  
{\displaystyle{\frac{h}{e}}}$ as given according to the flux  
quantization for electrons moving in the IQHE edge current on a ring  
with radius and width both equal to $l_B$. Then one obtains with  
the given $l_B$ according to the data in Ref. \cite{DK} for $ A =  
{\displaystyle{\frac{\hbar}{e}}}l^{-1}_B$ a value about $100 \mu m$  
for $A$, which is the mentioned observed legth for potential drops  
\cite{DK} \cite{dim}.

This result show that in view of the definition of magnetic length  
the measured value of $100 \mu m$ is a fundamental value for IQHE  
experiments on those samples \cite{DK} independent of other sample  
parameters.

\medskip
The fact that in other experiments where the electronic  
consentration is almost the same as in Ref. \cite{DK} but the  
filling factor is $\nu =4$, one observed a potential penetration of  
$\approx 70 \mu m$ \cite{font} is in good agreement with our  
theoretical results. Since for the $\nu =4$ filling factor one  
obtains according the given data in Ref. \cite{font} a magnetic  
length $l^{\prime}_B \approx 1.4 l_B \approx 1.4 \cdot 10^{-2}$ $\mu  
m$, where $l_B \approx 10^{-2}$ $\mu m$ is the magnetic length of   
samples in Ref. \cite{DK}.
Thus, the theoretical value of $ A =  
{\displaystyle{\frac{\hbar}{e}}}(l^{\prime}_B)^{-1}$ becomes  
$\approx 70 \mu m$ which is indeed the measured value according to  
the Ref. \cite{font} \cite{dim}.

\medskip
This circumstance explains why one observes such a distance from  
the boundary or edges in experiments concerning the IQHE \cite{DK}  
\cite{font}.

\medskip
Therefore, one should claim that the measured penetration length of  
electromagnetic potential on IQHE samples should depend, according  
to the theoretical value of $ A =  
{\displaystyle{\frac{\hbar}{e}}}(l_B)^{-1}$, only on the related  
value of $l^{-1}_B$ \cite{dim}.

\medskip
However, we shall mention further that the generality of this  
result requires a more subtle and theoretically more fundamental  
origin for this fact. Such an origin should be given, as it is  
mentioned already, by the quantum mechanical uncertainty- principle  
and relation, where a charged particle in presence of magnetic  
fields acquires a position uncertainty $\Delta X = l_B$.
Thus, considering $\Delta P = \Delta A = e A$, we are given under  
quantum mechanical conditions which apply to the QHE, the  
uncertainty relation $ eA\cdot l_B = \hbar$ which is the same  
relation as already mentioned. Here  
${\displaystyle{\frac{\hbar}{e}}}$ plays the same role in the  
quantum electrodynamical uncertainty as that played by $\hbar$ in  
the quantum mechanical uncertainty.

Therefore, in view of the fact that the value of  
${\displaystyle{\frac{\hbar}{e}}}$ is a  fixed quantity  
$\hbar^{\prime}$, the value of potential (drop) under IQHE  
conditions is always given by $A =  
{\displaystyle{\frac{\hbar^{\prime}}{l_B}}}$, no matter what other  
relevant quantities are.

Thus, in any IQHE sample one should measure for the potential drops  
on the edges the {\it related} value of $A =  
{\displaystyle{\frac{\hbar^{\prime}}{l_B}}}$ according to the value  
of $l_B$ from the experimental data of sample, as it is confirmed by  
the results in Ref. \cite {DK} and \cite{font} (see also  
\cite{dim}).

\medskip
In view of the fact that this is a result from the uncertainty  
principle and as such it is an invariant result, it depends only on  
the basics of "magnetic" quantization, i. e. on the uncertainty  
principle in quantum electrodynamics.

\medskip
Furthermore, it is expected that the observed length of the  
potential drop should be related with parameters of samples. This is  
indeed true, if one recalls that the concentration of charge  
carriers is indeed the main parameter of the sample and also the  
magnetic length depends on it.

\medskip
In conclusion let us mention that such a penetration length is also  
comparable with London's penetration length in superconductivity  
\cite{sup}.

\bigskip
Footnotes and references


\begin{thebibliography}{100}

\bibitem{all}
For a general review on QHE and its experimental setting see:

[1a] R.E. Prange and S.M. Girvin, ed., The quantum Hall effect,  
Graduate Texts in Contemporary Physics (Springer, New York, 1987);

[1b] A.H. Macdonald, ed., Quantum Hall effect: A Perspective,  
Perspectives in Condensed Matter Physics (Kluver Academic  
Publishers, 1989)

[1c] G. Morandi, The role of Topology in Classical and Quantum  
Physics, Lecture Notes in Physics m7 (Springer, New York 1992)

[1d] M. Janssen, et al, ed., J. Hajdu, Introduction to the Theory  
of the Integer Quantum Hall effect (VCH-verlag, Weinheim, New York,  
1994)

[1e] V. J. Emery (editor), Correlated Electron Systems, (World  
Scientific, Singapore 1993)

[1f] J. Froehlich, T. Kerler, Nuc. Phys. B354 (1991) 369-417.

[1g] A. Shapere and F. Wilczek (editors): Geometric Phases in  
Physics ( World Scientific, Singapore, 1989)


\bibitem{mein}
F. Ghaboussi, "On the Integer Quantum Hall Effect", KN-UNI-preprint-95-1;
"A Model of the Integer Quantum Hall Effect", KN-UNI-preprint-95-2,  
submitted for publication; See also "On the Hall-Effect and its  
Quantization", KN-UNI-preprint-95-3, submitted for publication.

\bibitem{kk}
K. von Klitzing, Physica B 204 (1995) 111-116;
R. Knott, W. Dietsche, K. von Klitzing, K. Eberl and K. Ploog,  
Semicond. Sci. Technol. 10 (1995) 117-126

\bibitem{DK}
W. Dietsche, K. v. Klitzing and K. Ploog, Potential Drops Across  
Quantum Hall Effect Samples- In the Bulk or Near the Edges? MPI fuer  
Festkoerperforschumg Stuttgart-preprint 1995;

\bibitem{nn}
According to the data about the IQHE samples in Ref. \cite{DK} the  
global concentration $n = 3.7 \cdot 10^{11} cm^{-2}$ and $\nu = 2$.  
Thus, one obtains $l_B \approx 10^{-2} \mu m$.

The measured pentration length is given to be about $100 \mu$ m  
which is almost exactly $|l^{-1}_B| \mu m$.

\bibitem{dim}
Recall that the mesured length should be considered according to  
the dimensinal structure where $\hbar$ contains $L^2$ dimensions  
according to its definition.


\bibitem{font}
P. F. Fontein, et al., Phys. Rev. B., 43, 12090 (1991). The given  
data in this report which are relevant for our calculation are $n =  
5.0 \cdot 10^{15}\, m^{-2}$ and $\nu = 4$.

\bibitem{sup}
It is well known that superconducting effects can be considered as  
to be related with the QHE: see Ref. [1f]; R. B. Laughlin: in Ref.  
[1g]; and A. Karlhede, et al.: in Ref. [1e]; See further for  
empirical confirmations: D. Jerome, in J. G. Bednorz, K. A. Mueller  
(Eds), Superconductivity,

( Spriger-Verlag, Berlin 1990).


\end{thebibliography}
\end{document}